\documentstyle[aps,prabib,preprint]{revtex}

\begin{document}

\draft
\title{Statistical Theory of Sedimentation of Disordered Suspensions}
\author{Hisao Hayakawa$^{1,2}$\thanks{e-mail: hisao@engels.physics.uiuc.edu}
and Kengo Ichiki$^1$\thanks{e-mail: ichiki@cmpt01.phys.tohoku.ac.jp}
}
\address{$^1$ Department of Physics, Tohoku University, Sendai 980-77, Japan
and \\
 $^2$ Department of Physics, University of Illinois at Urbana-Champaign, 1110
West Green Street, Urbana, IL 61801-3080, USA\thanks{Present address}}
\date{January 13, 1995}
\maketitle
\begin{abstract}
An analytical treatment for
the sedimentation rate of disordered suspensions is presented
in the context of a resistance problem.
 From the calculation it is confirmed that the lubrication  effect is important
in contrast to the suggestion by Brady and Durlofsky (Phys.Fluids {\bf 31}, 717
(1988)).
 The calculated
sedimentation rate
agrees well with the experimental results in all range of
the volume fraction.
\end{abstract}


The sedimentation of disordered suspensions is  important
in both technology and laboratory\cite{russel}.
The role of the sedimentation  is relevant to the current topics of statistical
mechanics such as
fluidized beds of gas-solid or liquid-solid mixtures
\cite{batchelor88,hayakawa,crighton},
 and
the density waves in the granular flows in vertical
tubes\cite{lee1994}.
 We believe the subject is a fundamental one
in fluid mechanics\cite{Kim}.
The rate of sedimentation for disordered suspensions
under gravity has yet to be determined theoretically
 except for a problem for  dilute spheres
with hard core interactions at a small Reynolds number
 \cite{russel,batchelor1972a}.

Our present understanding of theoretical studies
of sedimentation of monodisperse random suspensions
can be summarized as follows.
Batchelor\cite{batchelor1972a} has calculated the sedimentation rate in the
dilute limit of hard core particles with the radius $a$ based on the following
assumptions:
(i) The rate can be obtained from the combination  of the mobility matrix of
two particles and the two-body correlation function $g_{eq}(r)$  where $r$ is
the relative distance of particles, and (ii) the correlation function is
assumed to be $g_{eq}(r)=\theta(r-2a)$ , where $\theta(x)$ is the step function
$\theta(x)=1$ for $\ge 0$ and $\theta(x)=0$ otherwise. His result  at
 the volume fraction $\phi$ can be  written as
 $U(\phi)/U_0=1-6.55\phi+O(\phi^2)$ for $\phi\to 0$, where $U(\phi)$ the
sedimentation velocity at $\phi$ and $U_0$ is
 the equilibrium sedimentation velocity of one particle.
The result of Batchelor consists of two parts:
one is $1-5\phi$  from the Rotne-Prager tensor
which represents the effects of long-range hydrodynamic interaction,
and another $-1.55 \phi$  from the lubrication, the hydrodynamic repulsive,
force.
Extensions of this dilute theory
to concentrated suspensions require the account
of many body hydrodynamic interactions.
A generalization\cite{glendinning1982}, based on the method
of O'Brien\cite{obrien1979} predicts negative sedimentation rate for
$\phi>0.27$.
Brady and Durlofsky\cite{brady1988c}  have also obtained a negative
sedimentation rate
for $\phi> 0.23$ when they adopt well accepted correlation function $g_{eq}(r)$
for concentrated suspensions.
 As a result, they claim that the Rotne-Prager approximation
actually captures the correct features of sedimentation and ignored all of the
contributions from the lubrication force.
We feel, however, the statement by Brady and Durlofsky
\cite{brady1988c}
 unacceptable, because
there is no reason to ignore lubrication effects in the dilute
limit\cite{batchelor1972a}.
On the other hand,  Beenakker and Mazur\cite{beenakker1984a,mazur1982b}
 also calculated
the sedimentation rate based on an effective medium approximation and multipole
expansions.
Although
 they did not present an explicit expression of the sedimentation
rate,
  Ladd\cite{ladd1990} indicated that their result is better
than the result by Brady and Durlofsky\cite{brady1988c} for
concentrated suspensions.
In this  Rapid Communication, we wish to  demonstrate the relevance of
the lubrication force and improve the theory by
Brady and Durlofsky\cite{brady1988c}.
We also clarify the relationship between our theory and that by Beenakker and
Mazur\cite{brady1988c,beenakker1984a}.

The problem of sedimentation of $N$ particles with the radius $a$ at low
Reynolds numbers is
 equivalent to obtaining the resistance matrix ${\sf R}$ or
the mobility matrix ${\sf M}$ in
\begin{equation}
  \label{mob&res}
  {\bf U}
  =
  {1\over 6\pi\mu a}
  {\sf M}\cdot {\bf F}, \qquad {\sf M}={\sf R}^{-1},
\end{equation}
where ${\bf U}$ and ${\bf F}$ denote the sets of
the velocity field of $N$
particles and
the force exerted on $N$ particles, respectively, and $\mu$ is the shear
viscosity.
These mobility and resistance problems are not easy to solve
even numerically. One of the most successful numerical methods,
the Stokesian dynamics,
has been developed by Brady and his coworkers
\cite{brady1988a,Bossis}. The extension by Ladd\cite{ladd1990} also follows
a similar algorithm to the Stokesian dynamics.
They decouple the resistance matrix into the far-field part
$({\sf M}^{\infty})^{-1}$ and
the lubrication part ${\sf R}^{lub}$ as
\begin{equation}
\label{resistance}
  {\sf R}
  =
  \left({\sf M}^\infty\right)^{-1}
  +{\sf R}^{lub} ,
\end{equation}
where ${\sf R}^{lub}$ is calculated
 by the  pairwise additive expression of
the two-body lubrication matrix
$ {\sf R}^{lub}_{2B}
  =
  {\sf R}_{2B}
  -({\sf M}^\infty_{2B})^{-1}
$.
The resistance matrix is calculated as a function of the particle
configuration at each
numerical step. Then the force exerted on spheres and consequently the equation
of motion are obtained.
The success in the Stokesian dynamics suggests
that the problem for sedimentations should be considered based on a resistance
picture. In fact, some unphysical results of simulations
 based on a mobility picture supports
this statement.
We  may understand the relevance of a resistance picture as follows.
Since the contribution of the lubrication is  proportional
to the number of particles, as will be shown,
 the direct addition of the lubrication for the mobility
 cannot avoid a negative sedimentation rate.
In other words, the linear contribution of the lubrication to the drag is
reasonable,
 while the linear addition
of the lubrication to the mobility cannot produce any nonlinear
complicated motion of particles in experiments.

Thus, we are not surprized by the failure of direct generalizations
 of Batchelor's theory which is described as a mobility problem.
We must calculate the sedimentation rate in the context of a resistance
problem.
The problem is, thus,  reduced to obtaining
$  <\left({\sf M}^\infty\right)^{-1}>
  +<{\sf R}^{lub}> , $
where the bracket is the  average over the particle configurations.
Note that $<({\sf M}^\infty)^{-1}>$ and $<{\sf R}^{lub}>$ are
the scalar quantities.
The far-field part can be calculated from
$<({\sf M^{\infty}})^{-1}>\simeq <{\sf M^{\infty}}>^{-1}=\tilde M(k=0)^{-1}$,
where $\tilde M(k)$  is defined by
\begin{equation}
 \label{M^infty}
 \tilde M(k) =
  1
  +n \int_V e^{i{\bf k}\cdot{\bf r}_{12}}(g_{eq}({\bf r}_{12})-1)
  \hat{{\bf k}}\cdot{\sf G}({\bf r}_{12})\cdot\hat{{\bf k}}
  d{\bf r}_{12}.
\end{equation}
Here
$\hat{{\bf k}}={\bf k}/k$, the relative
position ${\bf r}_{12}$ of the particles 1 and 2,
$n$ is the number density of
partcles.
The explicit representation of the Fourier component of the tensor ${\sf
G}=\{G_{ij}\}$
is given by \cite{mazur1982b,ladd1993}
\begin{equation}
  \label{connector}
  G_{ij}({\bf k})=6\pi a
  \frac{j_0(ka)^2}{ k^2}
  (\delta_{ij}-\frac{k_ik_j}{k^2})
\end{equation}
with the spherical Bessel function $j_0(ka)$.
For later discussion we drop the suffix of ${\bf r}_{12}$ and
assume the isotropy of systems as $g_{eq}(r=|{\bf r}|)$.

 The correlation function $g_{eq}(r)$ can be approximated
 \cite{Bossis,ladd1993} by
 the  equilibrium distribution function for
hard sphere systems based on the Percus-Yevick approximation\cite{percus1958}.
The Fourier transform of $g_{eq}(r)-1$,
$h(k)$ is represented by\cite{wertheim1963}
\begin{equation}
  \label{P-Y}
  h(k)
  =-\frac{4\pi a^3\tilde{c}(ka)}{1+3\phi \tilde{c}(ka)} ,
\end{equation}
where $\tilde c(x)$ is the direct correlation function which also depends on
$\phi$.
The correlation function in (\ref{P-Y})  reduces to $g_{eq}(r)=\theta(r-2a)$ in
the dilute limit.
 From (\ref{P-Y})
 we can evaluate $<{\sf M^{\infty}}>=\frac{2}{\pi}\int_0^{\infty}dx(\frac{{\rm
sin}x }{x})^2(1+3\phi{\tilde c}(x))^{-1}$ numerically.
Brady and Durlofsky\cite{brady1988c}
 evaluated this\cite{Note} by using the Laplace transform
of the Percus-Yevick distribution  function\cite{Mansoori}
 and the method
of O'Brien\cite{obrien1979} as
\begin{equation}
  <{\sf M}^\infty>
 \simeq
    {(1-\phi)^3\over(1+2\phi)} \quad ,
\label{Rotne-Prager}
\end{equation}
which is a correct evaluation of the contribution from the far-field part.

Now, we evaluate the contribution from $<{\sf R}^{lub}>$.
For simplicity of the argument, we neglect contributions from higher order
moments such as torque and shear.
Since $<{\sf R}^{lub}>$ is evaluated from a pairwise  additive
approximation, $<{\sf R}^{lub}>$ is represented by
\begin{equation}
  <{\sf R}^{lub}>
  =
  n\int_V d{\bf r}
  g_{eq}(r)\hat{{\bf k}}\cdot\left[
  {\sf A}_{11}+{\sf A}_{12}
  -\left\{ {( {\sf M}^\infty_{2B} )^{-1}}_{11}
  +{ ({\sf M}^\infty_{2B})^{-1}}_{12} \right\}\right]\cdot
  \hat{{\bf k}} .
 \label{R^{lub}}
\end{equation}
The tensor
${\sf A}_{\alpha\beta}$ is a part of ${\sf R}_{2B}$ and its sufficies represent
the particles.
The tensor ${\sf A}_{11}+{\sf A}_{12}$, thus, is given by
\begin{equation}
\label{A_{ij}}
{\sf A}_{11}+{\sf A}_{12}=\pmatrix{
Y_{11}+Y_{12} & 0 & 0\cr
0 & Y_{11}+Y_{12} & 0 \cr
0 & 0 & X_{11}+X_{12} \cr } ,
\end{equation}
 where the explicit representations of $X_{ij}$ and $Y_{ij}$ are given by
Jeffrey and Onishi
\cite{jeffrey1984} as a series expression.
On the other hand,  $({\sf M}^\infty_{2B})_{11}$  is the unit  tensor
and $({\sf M}^\infty_{2B})_{12}$ is the Rotne-Prager tensor which is
represented  by
\begin{equation}
  \label{M_{12}}
  ({\sf M}^\infty_{2B})_{12}=x^{\infty}(r)
  \hat{{\bf r}}\hat{{\bf r}}
  +y^{\infty}(r)\left({\sf I}-\hat{{\bf r}}\hat{{\bf r}}\right) ,
\end{equation}
where $x^{\infty}(r)=(3/2)(r/a)^{-1}-(r/a)^{-3}$ and
$y^{\infty}(r)=(3/4)(r/a)^{-1}+(1/2)(r/a)^{-3}$.
The tensor
${({\sf M}^\infty_{2B})^{-1}}_{11}
+{({\sf M}^\infty_{2B})^{-1}}_{12}$
can be  readily calculated as
\begin{equation}
  \label{resi}
  {({\sf M}^\infty_{2B})^{-1}}_{11}
  +{({\sf M}^\infty_{2B})^{-1}}_{12} =
  X^\infty(r)\hat{{\bf r}}\hat{{\bf r}}
  +Y^\infty(r)\left({\sf I}-\hat{{\bf r}}\hat{{\bf r}}\right),
\end{equation}
where   $X^{\infty}(r)
=
   (1+x^\infty(r))^{-1}$ and
$  Y^{\infty}(r)
  =(1+y^\infty(r))^{-1}$.
Thus, the average of the contribution from the lubrication part
is described by
\begin{equation}
 \label{R^{lub}-f}
 <{\sf R}^{lub}>
  =
  \phi\int_2^{\infty} dz\
  z^2g_{eq}(r) W(z)  \quad ,
\end{equation}
where $z=r/a$ and
\begin{equation}
\label{W(z)}
  W(z)=X_{11}+X_{12}
  +2Y_{11}+2Y_{12}
  -{6z^3(-2+5z^2+4z^3)\over(-2+3z^2+2z^3)(2+3z^2+4z^3)}.
\end{equation}
With the aid of the exact result by Jeffrey and Onishi\cite{jeffrey1984}
$W(z)$ can be evaluated as
\begin{equation}
 W(z) =
  {21\over 4}{1\over z^4}
  -{789\over 64}{1\over z^5}
  +O(\frac{1}{z^6})
\end{equation}
for $z\gg 1$. Thus we can evaluate $<{\sf R}^{lub}>$
 by the numerical integral.
For the practical purpose, it is convenient to have an explicit expression for
$<{\sf R}^{lub}>$.
If we assume $g_{eq}(r)=\theta(r-2a)$, $<{\sf R}^{lub}>$ is approximately
represented by
\begin{equation}\label{evaluate}
<{\sf R}^{lub}>
\simeq
  \phi\displaystyle\int_{2}^{20} dz
  z^2W(z)
  +\phi\displaystyle\int_{20}^\infty dz
  z^2
  \left(
  \displaystyle\frac{21}{4}\displaystyle\frac{1}{z^4}
  -\displaystyle\frac{789}{64}\displaystyle\frac{1}{z^5}\right)
\simeq  1.492\phi
\end{equation}
When we compare the result (\ref{evaluate}) with the one obtained with the aid
of  the Percus-Yevick distribution function for $g_{eq}(r)$,
we find that the two results have no significant difference (see Fig.1).
This statement is applicable to the calculation for the lubrication part of
the mobility matrix as $<{\sf M}^{lub}>\simeq -1.55\phi$.
We thus confirm that the contribution from the lubrication is insensitive to
the form of $g_{eq}(r)$ and is proportional to $\phi$.
Thus we should solve the problem in the context of a resistance problem
to avoid a negative sedimentation rate.

 From (\ref{Rotne-Prager}) and (\ref{evaluate}) we obtain
\begin{equation}
 \frac{U}{U_0}
  \simeq
  \frac{1}{<{\sf M}^\infty>^{-1}+<{\sf R}^{lub}>}
  =
{(1-\phi)^3\over{1+2\phi+1.492\phi(1-\phi)^3}} .
 \label{result}
\end{equation}
As will be shown, this result is sufficiently close to experimental values.
 The dilute limit of our result $  U/U_0=1-6.49\phi+O(\phi^2) $
is slightly different from Batchelor's result
$U/U_0=1-6.55 \phi+O(\phi^2)$.
 This discrepancy comes from the relation
${\sf R}^{lub}\ne ({\sf M}^{lub})^{-1}$.
The true dilute limit should be calculated under the considerations of
all of higher order moments\cite{batcom}.
 It is worthwhile, however,  to
indicate that our theory essentially
resolves the contradiction about contributions from the lubrication
in the result by Brady and Durlofsky\cite{brady1988c}.

Now we compare our result with that by
Beenakker and Mazur \cite{beenakker1984a}.
They rewrite the renormalized (4) as
\begin{equation}
\label{B-M84}
 \tilde M_{\gamma_0}(k) =
  1
  +\hat{{\bf k}}\cdot{\sf G}_{\gamma_0}({\bf r}=0)\cdot\hat{{\bf k}}
  +n\int d{\bf r}\
  e^{i{\bf k}\cdot{\bf r}}
  \hat{{\bf k}}\cdot{\sf G}_{\gamma_0}({\bf r})\cdot\hat{{\bf k}}
  \left\{g_{eq}(r)-1\right\}
  \label{eq:g(k)d(k)},
\end{equation}
where  ${\sf G}_{\gamma_0}({\bf r})$ is given by
\begin{equation}
{\sf G}_{\gamma_0}(r)={\tilde {\sf G}}(r)-\int \frac{d{\bf k}}{(2\pi)^3}e^{i
{\bf k}\cdot {\bf r}}\frac{\phi S_{\gamma_0}(ka)}{1+\phi S_{\gamma_0}(ka)}{\sf
G}(k).
\end{equation}
Here $S_{\gamma_0}(x)$ is the structure factor and  $\tilde {\sf G}(r)=0$ for
$r=0$ and $\tilde {\sf G}(r)={\sf G}(r)$ for $r\ne 0$.
Substituting (\ref{P-Y}) into
(\ref{B-M84}) and noting $<{\sf M}>=\tilde M_{\gamma_0}(k=0)$, we obtain
$  <{\sf M}>
  =
{2\over \pi}
 \int_0^\infty dx\
  \left({\sin x\over x}\right)^2
  \{(1+\phi S_{\gamma_0}(x))
  ( 1+3\phi\tilde{c}(x))\}^{-1}
$ .
The function $S_{\gamma_0}(x) $
tends to $5/2$
for dilute case and small $x$. In the dilute limit, the result by
Beenakker and Mazur\cite{beenakker1984a} is reduced to
$ U/U_0\simeq 1-(15/2) \phi +O(\phi^2), $
which is considerably away  from Batchelor's result \cite{batchelor1972a}.
Even in concentrated cases, $S_{\gamma_0}(x)$ still may be replaced by
$5/2$, although its actual expression is complicated.
With the aid of (\ref{Rotne-Prager}) an approximate expression
of Beenakker and Mazur\cite{beenakker1984a} is given by
\begin{equation}
\label{c-BM}
\frac{U}{U_0}\simeq \frac{(1-\phi)^3}{(1+2\phi)(1+5\phi/2)} .
\end{equation}
 From (\ref{c-BM}), it is easy to understand that  Beenakker
and Mazur\cite{beenakker1984a}
 renormalize the Rotne-Prager tensor by taking into account the contribution
from the structure factor.
The deviation from Batchelor's result in the dilute limit
suggests  that they miss the quantitative description for the short-range
force,
because their effective field approximation
 includes only parts of the lubrication by a collection of ladder diagrams.
  Their theory, however,  may be good
for dense suspensions where the requirement
for their approximation may be satisfied.

Let us compare theoretical results with experimental ones
\cite{buscall1982,bacri1986,paulin1990,xue1992,dekruif1987}(Fig.2).
We  recognize that our theory improves the result
by Brady and Durlofsky \cite{brady1988c} and achieves good agreement with
experiments.
 Therefore, we conclude that the contribution from
the lubrication force is small but relevant.  For $\phi<0.2$,
it seems that our result is better than that by Beenakker and
Mazur\cite{beenakker1984a}.
 In high concentration regions, however, our sedimentation rate
is a little larger than the experimental values, while the prediction by
Beenakker and Mazur\cite{beenakker1984a} works well.
This disagreement between our theory and experiments
in concentrated regions seems to come from
the neglect of higher order moments.
The high sedimentation rate without higher order moments for a regular
configuration
of particles has been reported\cite{brady1988a}.
To check this tendency for random particle configurations we have performed a
simulation for 50 particles
based on the Stokesian dynamics, where we neglect the contributions from higher
order multipole expansions.
In our simulation
the particle configuration is at random
and average 100 configurations for each $\phi$ to calculate the sedimentation
rate.
When we neglect the statistical error, the tendency of high sedimentation rates
in large $\phi$
 coincides with that of our theory.

In conclusion, we have confirmed that the calculation of sedimentation rate
should
be performed in the context of a resistance problem.
It is not surprising that
the direct generalization of Batchelor's theory
based on the mobility picture gives us wrong answers.
Thus, we should include the lubrication effects in contrast to the claim by
Brady and Durlofsky\cite{brady1988c}.
 Our method including
the lubrication force is  an adequate systematic approach
to extend the dilute theory.
 We demonstrate that the lowest order contribution to the sedimentation rate of
the lubrication force becomes closer
to experimental values than that by the Rotne-Prager approximation.
 The discrepancy between our  calculation and
experiments at high $\phi$ should be improved if we include the contribution
from torque and other moments. The consecutive improvement of our calculation
of
the sedimentation rate
will be reported elsewhere.

\par
\vspace*{0.4cm}
\par
We thank T.Ohta and S.Sasa for stimulating discussion and Y.Oono for his
critical reading and his useful comments.
This work is, in part, supported by Foundation for
Promotion of Industrial Science and by National Science Foundation Grant No.
NSF-DMR-93-14938.


\begin{references}
\bibitem{Note}
The equivalency between this expression and O'Brien's method is
not trivial.
We , however, have checked that the deviation between the direct
integration of the Percus-Yevick distribution function and
(\ref{Rotne-Prager}) is very small.
\bibitem{batcom}
There is no reason to believe that Batchelor's result is exact, since
 the correct calculation should be based on the picture of a resistance
problem.
\end{references}

\begin{figure}
\caption{
The comparison of several theoretical and numerical
predictions of the sedimentation
rate $U(\phi)/U_0$ as  functions of $\phi$. For
Eq.(11) with PY,
we use the Percus-Yevick distribution function for $g_{eq}(r)$
in (11) to evaluate $<{\sf R}^{lub}>$. The result of
Ref.\protect\cite{ladd1990} is obtained from his
precise simulation. Ref.\protect\cite{beenakker1984a} is from their
Fig.2 with $k=0$ and its approximate expression is given by
(\protect\ref{c-BM}).
}
\label{fig:1}
\end{figure}

\begin{figure}
\caption{
The comparison of several theoretical results
with experimental results for $U(\phi)/U_0$.
We also plot the data of our Monte-Carlo simulation. See the text
for the details.
}
\label{fig:2}
\end{figure}



%
%

%
%

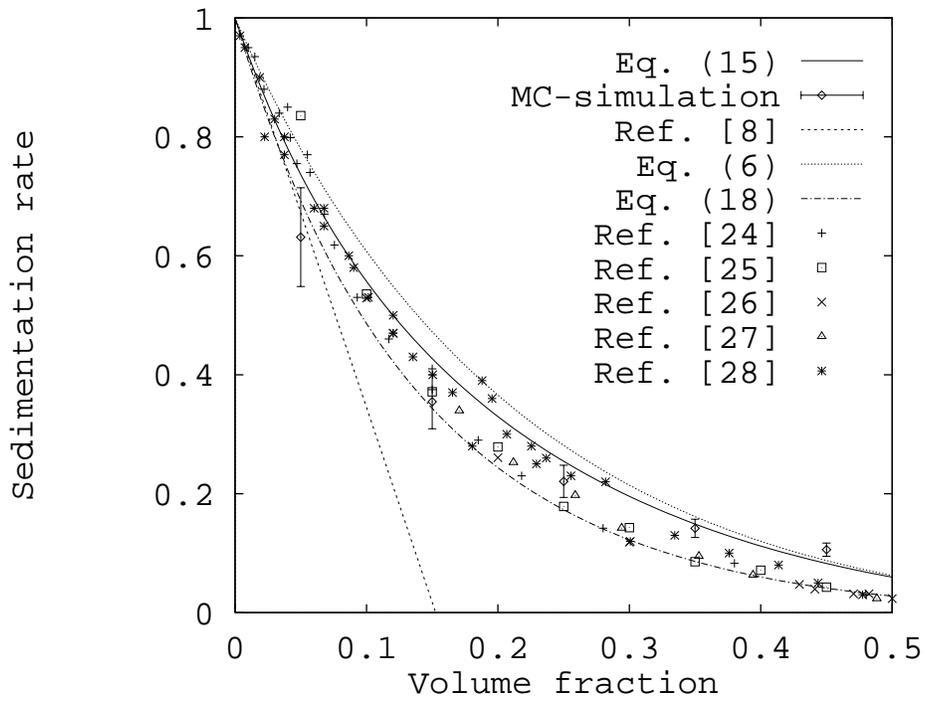
\end{document}